\documentclass[prb,twocolumn,showpacs,superscriptaddress,preprintnumbers,floatfix,amsmath,amssymb]{revtex4-1}

\usepackage{amssymb}
\usepackage{graphicx}
\usepackage{dcolumn}
\usepackage{bm}
\usepackage{color}
\begin{document}

\title{Effect of impurity scattering on superconductivity in K$_2$Cr$_3$As$_3$}

\author{Y. Liu}
\affiliation{Department of Physics, Zhejiang University, Hangzhou 310027, China}
\affiliation{Collaborative Innovation Centre of Advanced Microstructures, Nanjing 210093, China}
\author{J. K. Bao}
\affiliation{Department of Physics, Zhejiang University, Hangzhou 310027, China}
\affiliation{Collaborative Innovation Centre of Advanced Microstructures, Nanjing 210093, China}

\author{H. K. Zuo}
\affiliation{Wuhan National High Magnetic Field Center, School of Physics,
Huazhong University of Science and Technology, Wuhan 430074, China}

\author{A. Ablimit}
\affiliation{Department of Physics, Zhejiang University, Hangzhou 310027, China}
\affiliation{Collaborative Innovation Centre of Advanced Microstructures, Nanjing 210093, China}v
\author{Z. T. Tang}
\affiliation{Department of Physics, Zhejiang University, Hangzhou 310027, China}
\affiliation{Collaborative Innovation Centre of Advanced Microstructures, Nanjing 210093, China}
\author{C. M. Feng}
\affiliation{Department of Physics, Zhejiang University, Hangzhou
310027, China}
\author{Z. W. Zhu}
\affiliation{Wuhan National High Magnetic Field Center, School of Physics,
Huazhong University of Science and Technology, Wuhan 430074, China}

\author{G. H. Cao}\email[]{ghcao@zju.edu.cn}
\affiliation{Department of Physics, Zhejiang University, Hangzhou
310027, China}
\affiliation{Collaborative Innovation Centre of Advanced Microstructures, Nanjing 210093, China}
\affiliation{State Key Lab of Silicon Materials, Zhejiang University, Hangzhou 310027, China}

\date{\today}

\begin{abstract}
Impurity scattering in a superconductor may serve as an important probe for the nature of superconducting pairing state. Here we report the impurity effect on superconducting transition temperature $T_\mathrm{c}$ in the newly discovered Cr-based superconductor K$_2$Cr$_3$As$_3$. The resistivity measurements show that the crystals prepared using high-purity Cr metal ($\geq$99.99\%) have an electron mean free path much larger than the superconducting coherence length. For the crystals prepared using impure Cr that contains various \emph{nonmagnetic} impurities, however, the $T_\mathrm{c}$ decreases significantly, in accordance with the generalized Abrikosov-Gor'kov pair-breaking theory. This finding supports a non-$s$-wave superconductivity in K$_2$Cr$_3$As$_3$.
\end{abstract}


\maketitle

The discovery of superconductivity in alkali-metal chromium arsenides $A_2$Cr$_3$As$_3$ ($A$=K, Rb, Cs)\cite{bao,Rb233,Cs233} has drawn much attention\cite{imai,zgq,yhq1,yhq2,musr,sll,raman,zzw,canfield1,canfield2,jp,caoc,hujp1,hujp2,zy,dai,hujp3,alemany,hujp4} primarily because of the possibly exotic superconductivity. The new superconducting family is structurally characterized by the infinite [(Cr$_3$As$_3$)$^{2-}$]$_{\infty}$ linear chains, which bears a quasi-one-dimensional characteristic. The superconducting transition temperature, $T_{\mathrm{c}}$, is 6.1 K, 4.8 K and 2.2 K for $A$=K, Rb and Cs, respectively. Unconventional superconductivity is evidenced by a growing body of experimental observations.\cite{bao,Rb233,imai,zgq,yhq1,yhq2,musr,sll,raman,zzw} For example, the K$_2$Cr$_3$As$_3$ superconductor shows an unusually large upper critical field, $H_{\mathrm{c2}}$, which exceeds the Bardeen-Cooper-Schrieffer (BCS) weak-coupling Pauli limit by a factor of three.\cite{bao,canfield1,canfield2,jp} The angular dependence of $H_{\mathrm{c2}}$ further reveals a \emph{fully} anisotropic Pauli-limiting behavior due to the spin-orbit coupling, indicating a dominant spin-triplet pairing.\cite{zzw} The triplet superconductivity is supported by the ferromagnetic spin fluctuations, first suggested by theoretical calculations,\cite{caoc,hujp1} then verified by the nuclear magnetic resonance (NMR) study\cite{zgq}. Furthermore, the theoretical modelling\cite{hujp2,zy,dai,hujp3} consistently favors the spin-triplet pairing channel with a possible $p_z$-wave\cite{hujp2} or $f$-wave\cite{zy} pairing symmetry.

Nevertheless, the triplet superconductivity seems to be inconsistent with the preliminary observation of insensitivity of $T_\mathrm{c}$ for the K$_2$Cr$_3$As$_3$ samples with different residual resistivity.\cite{canfield1,canfield2} As is known, $T_\mathrm{c}$ hardly changes by nonmagnetic scattering for an $s$-wave spin-singlet superconductor, according to the Anderson¡¯s theorem.\cite{anderson,rmp} For a non $s$-wave superconductor in which the superconducting gap function changes sign around the Fermi surface, however, the nonmagnetic scattering breaks the Cooper pairs,\cite{larkin,millis,radtke,rmp} just in a manner of the pair breaking by magnetic impurities in a conventional $s$-wave superconductor.\cite{AG} Indeed, severe suppression of $T_\mathrm{c}$ by nonmagnetic scattering have been observed in many unconventional superconductors including the $f$-wave superconductor UPt$_3$\cite{UPt3}, the $p$-wave superconductor Sr$_2$RuO$_4$\cite{Ru} as well as the $d$-wave superconducting cuprates\cite{cuprate}. That is why the insensitivity of $T_\mathrm{c}$ to nonmagnetic impurities was argued to oppose a triplet pairing in K$_2$Cr$_3$As$_3$.\cite{canfield1,canfield2}

We note that the $H_{\mathrm{c2}}$ values measured are very large.\cite{canfield2,zzw} This means that the superconducting coherence lengths at zero temperature, $\xi_{\bot}(0)$ and $\xi_{\parallel}(0)$, are relatively short. By using the Ginzburg-Landau (GL) relations, $H_{\mathrm{c2},\parallel}^{\mathrm{orb}}(0)=\Phi_{0}/[2\pi \xi_{\bot}(0)^2$] and $H_{\mathrm{c2},\bot}^{\mathrm{orb}}(0)=\Phi_{0}/[2\pi \xi_{\bot}(0) \xi_{\parallel}(0)$], where $\Phi_{0}$ is the flux quantum, $\xi_{\bot}(0)$ and $\xi_{\parallel}(0)$ are estimated to be 2.53 and 3.5 nm, respectively.\cite{zzw} Therefore, one may expect a weak depression of $T_\mathrm{c}$ if the electron mean free path along the $c$ direction [(Eq. (1)], $\ell_{\parallel}$, is much larger than the superconducting coherence length $\xi_{\parallel}(0)$ (here we consider that superconductivity originates in the infinite [(Cr$_3$As$_3$)$^{2-}$]$_{\infty}$ chains), according to the Abrikosov-Gor'kov (AG) equation.\cite{AG} This motivated us to systematically study the effect of impurity scattering on $T_\mathrm{c}$ in K$_2$Cr$_3$As$_3$ single crystals. We initially tried to dope nonmagnetic Zn ions, unfortunately, the Zn solubility seems to be vanishingly small. Later we succeeded in changing the residual resistivity by using different-purity Cr sources. As a result, we found that the impurity scattering does suppress the $T_\mathrm{c}$, which basically obeys the AG equation. This finding solves the difficulty of triplet superconductivity with regard to the impurity scattering.

\section{Materials and measurements}\label{sec:2}
\paragraph{Crystal growth} Single crystals of K$_2$Cr$_3$As$_3$ were grown by a self-flux method using the starting materials (Alfa Aesar) of elements K pieces (99.95\%), Cr and As powder (99.999\%).\cite{bao} The main impurity in the K pieces is Na ($\sim$20 ppm), which is expected to substitute for K only, and thus it does not induce any disorder in the superconducting-actively [(Cr$_3$As$_3$)$^{2-}$]$_{\infty}$ chains. To alter the impurity concentrations at the Cr sites, we used different Cr sources: chromium powders [99\% (the resulted crystals are labelled Sample \#4), 99.95\% (Sample \#3) and 99.995\% (Sample \#2)] and chromium crystallites [99.995\% (Sample \#1)]. For the 99\% Cr powder (the oxygen content is 0.83\%), the sum of impurity metals (SIM) is 0.11\% with the main impurities of 0.09\%Fe and 0.0048\%Al. For the 99.95\% Cr powder, the SIM is 0.008\% with the main impurities of 0.003\%Fe, 0.001\%Ag, 0.0009\%Ga and 0.0008\%V. For the 99.995\% Cr powder and crystals, the SIM is 0.0024\% and 0.0019\%, respectively.

We employed simplified procedures for the crystal growth. Mixtures of K pieces, Cr metal and As pieces in a molar ratio of 6:1:7 were first loaded in an alumina crucible. The crucible was then jacketed in a Ta tube welded, followed by sealing in an evacuated quartz ampoule. The sample-loaded ampoule was heated slowly in a muffle furnace to 1273 K, holding for 12-24 h, and cooled down to 1123 K in 5 h. Single crystals are expected to grow up when further cooling down to 973 K at a rate of 2 K/h. The as-grown crystals are shiny, rod-like, and black in color, with a typical size of 2$\times$0.05$\times$0.05 mm$^3$. They are extremely air-sensitive, and any exposure to air should be avoided as far as possible (we employed an argon-filled glove box with the water and oxygen content below 0.1 ppm).

\paragraph{Electrical resistivity measurements} The electrical resistivity ($\rho_{\parallel}$) with the electric current flowing along the rod direction was measured using a standard four-terminal method. Special care should be taken to avoid the sample's degradation into KCr$_3$As$_3$, which is not superconducting.\cite{bao2} Various kind of electrode contacts were tried to prevent even slight damage of samples, and the dense silver paste (DuPont 4929N) was found to be an optimal contact medium (note that the electrodes were made in the glove box with a minimized amount of oxygen and water). It is more challenging to accurately measure the absolute resistivity owing to the small size and easy cleavage of crystals along the rod direction. To minimize the measurement errors, the electrodes for the voltage measurement were separated as far as possible (see the inset of figure 1), and in this circumstance, the uncertainty of the absolute $\rho$ value is mainly originated from the determination of the sample cross-section area ($\sim$30\%). Nonetheless, the room-temperature resistivity ($\rho_{\mathrm{rt}}$) measured scatters from 100 to 500 $\mu \Omega$ cm, independent of the purity of the Cr source. In general, $\rho_{\mathrm{rt}}$ changes insignificantly with the impurity scattering, as is exemplified in Sr$_2$RuO$_4$ crystals where $\rho_{\mathrm{rt}}$ is measured to be 121$\pm$3 $\mu \Omega$ cm, independent of the residual resistivity.\cite{Ru} We also note that the possible sample cleavage and degradation both lead to an overestimation of the resistivity, which might explain the relatively large $\rho_{\mathrm{rt}}$ values ($\sim$1000 $\mu \Omega$ cm) in previous reports.\cite{canfield1,jp} This means that the low values of $\rho_{\mathrm{rt}}$ measured should reflect the intrinsic property. Therefore, the scope of $\rho_{\mathrm{rt}}$ is reasonably determined to be 150$\pm$50 $\mu \Omega$ cm.

\section{Results and discussion}\label{sec:3}

\begin{figure*}
\centering
\includegraphics[scale=1.0]{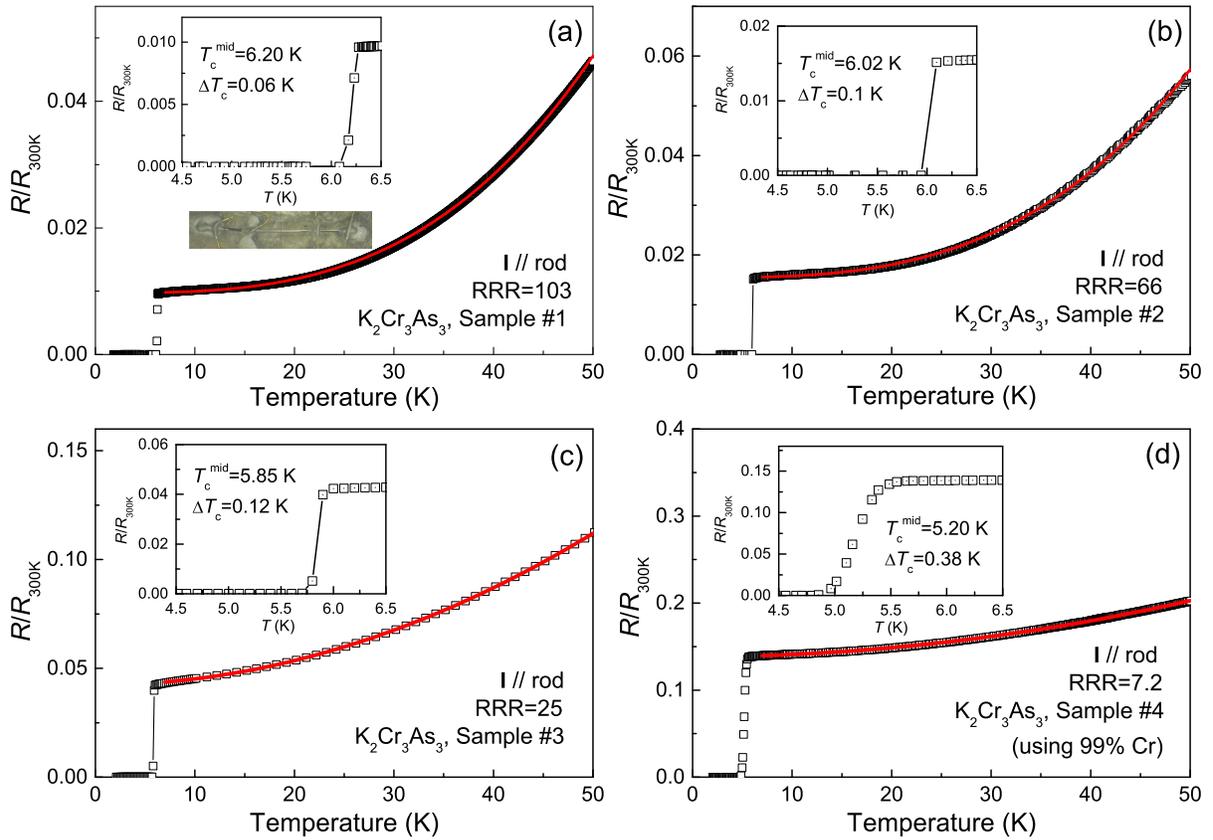}
\caption{Temperature dependence of resistivity with electric current flowing along the rod direction of different K$_{2}$Cr$_{3}$As$_{3}$ crystals. A normalized scale, $R/R_{\mathrm{300K}}$, is used for the resistivity axes. Shown in (a), (b), (c) and (d) are some typical $R(T)$ data in which Samples $\#$1, $\#$2, $\#$3 and $\#$4 are taken from four different batches prepared using different-purity Cr sources (see details in the text), respectively. The red lines are fitted curves with a power law. The insets zoom in the superconducting transition from which the superconducting transition temperature $T_{\mathrm{c}}^{\mathrm{mid}}$ (the 50\% point of the transition) and the transition width $\Delta T_{\mathrm{c}}$ are obtained. The lower inset of (a) is the photo of the sample that is connected with four gold wires ($\phi$=30 $\mu$m) by silver pastes.}
\label{fig1}
\end{figure*}

Figure \ref{fig1} shows the $R(T)$ data for different K$_{2}$Cr$_{3}$As$_{3}$ crystals. The high-temperature (50 K$<T<300$ K) $R(T)$ data are essentially linear (not shown here), similar to the previous report.\cite{canfield1} Since the $\rho_{\mathrm{rt}}$ values turn out to be independent of the samples within the measurement uncertainty, we employ a normalized scale, $R(T)/R_{\mathrm{300K}}$, for the resistivity axes. The $R(T)$ data in a temperature range of 7 K$\leq T\leq50$ K basically follow a power law, $R(T)=R_{0}+AR^{\alpha}$. The data fitting yields three parameters: the residual resistance $R_{0}$, the coefficient $A$ and the power $\alpha$.

We find that the crystals prepared using different-purity Cr sources yield distinct values of the above parameters. With using high-purity ($\geq$99.995\%) Cr, the residual resistance ratio (RRR), i. e. a ratio of $R_{\mathrm{300K}}$ and $R_{0}$, is always large (from 50 to 300). Simultaneously, the superconducting transitions are very sharp, as parameterized by the transition width $\Delta T_{\mathrm{c}}$ (the temperature difference between 10\% and 90\% values in the extrapolated normal-state resistance). The $T_{\mathrm{c}}$ value, defined by the transition midpoint $T_{\mathrm{c}}^{\mathrm{mid}}$ here, tends to increase with the RRR value, ranging from 6.0 K to 6.25 K. For Sample $\#3$, which is prepared using 99.95\% Cr, the RRR value becomes 25, and the $T_{\mathrm{c}}^{\mathrm{mid}}$ decreases to 5.85 K. The RRR value of Sample $\#4$ (prepared using 99\% Cr) is only 7.2, and the $T_{\mathrm{c}}^{\mathrm{mid}}$ drops to 5.20 K with an enhanced $\Delta T_{\mathrm{c}}$ of 0.38 K. The fitted $\alpha$ values tend to decrease with increasing impurities, ranging from 3.1 for Sample \#1 to 2.0 for Sample \#4. The $\alpha$ values for Samples \#1 and \#2 are consistent with the previous report.\cite{canfield1} The change in $\alpha$ may reflect a disorder effect in the electron-correlated quasi-one-dimensional system.

\begin{figure}
\centering
\includegraphics[width=8cm]{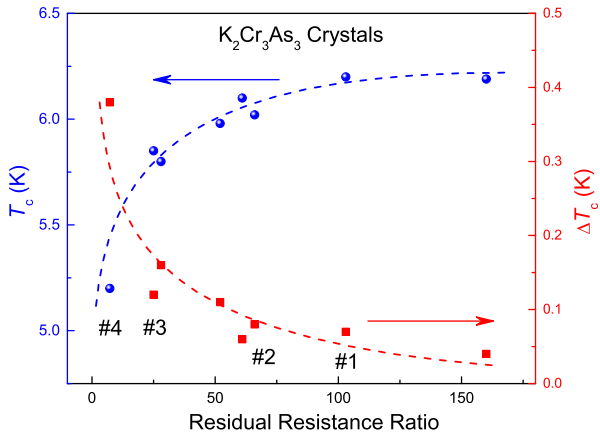}
\caption{(Color online) The superconducting transition temperature (left axis) and the superconducting transition width (right axis) as functions of the residual resistance ratio (RRR) for different K$_{2}$Cr$_{3}$As$_{3}$ crystals. The dashed lines are guides to the eye.}
\label{fig2}
\end{figure}

Although there are some uncertainties for the absolute residual resistivity, the RRR values are accurately measured for a given sample. Figure \ref{fig2} show the RRR dependence of $T_{\mathrm{c}}$ and $\Delta T_{\mathrm{c}}$ for the K$_{2}$Cr$_{3}$As$_{3}$ crystals. $T_{\mathrm{c}}$ tends to saturate at a large RRR value. When the RRR value is less than $\sim$25, $T_{\mathrm{c}}$ starts to drop rapidly, accompanied with a severe broadening in the superconducting transition. This phenomenon is reminiscent of the impurity effect on $T_{\mathrm{c}}$ in Sr$_2$RuO$_4$.\cite{Ru}

\begin{figure}
\centering
\includegraphics[scale=1.0]{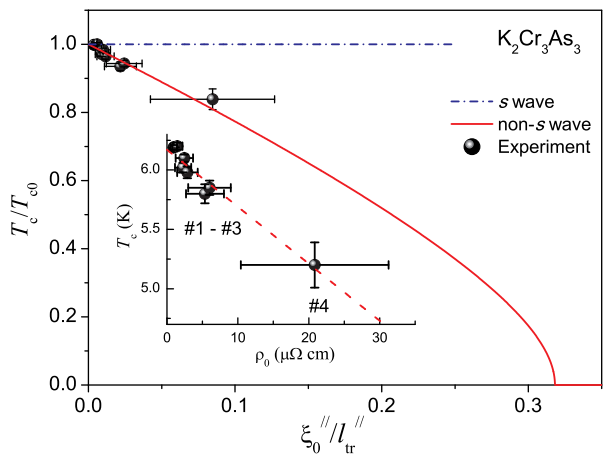}
\caption{(Color online) The reduced superconducting transition temperature, $T_{\mathrm{c}}/T_{\mathrm{c0}}$, as a function of the ratio of superconducting coherence length and electron mean free path along the $c$ axis, $\xi_{0}^{\parallel}/l_{\mathrm{tr}}^{\parallel}$, in K$_{2}$Cr$_{3}$As$_{3}$. The red solid line represents the result from the generalized Abrikosov-Gor'kov theory. The blue dash-dot line shows an $s$-wave behavior according to Anderson's theorem. The inset shows the relation between $T_{\mathrm{c}}$ and the residual resistivity $\rho_{0}$, in which the dashed line shows the linear fit. The horizontal error bars are based on the measurement uncertainty for $\rho_{\mathrm{rt}}$, and the vertical error bars denote the superconducting transition width.}
\label{fig3}
\end{figure}

As stated above, $\rho_{\mathrm{rt}}$ (150$\pm$50 $\mu \Omega$ cm) does not depend on the purity of the Cr source. This allows us to determine the residual resistivity $\rho_{0}$ from the RRR values. The inset of figure \ref{fig3} shows the relation between $T_\mathrm{c}$ and $\rho_{0}$. As shown, most data points are gathered around 6 K, which is consistent with the apparently weak sensitivity of $T_\mathrm{c}$ with $\rho_{0}$. By closer examinations, however, one may see that all the data points basically collapse on a straight line. Namely, $T_\mathrm{c}$ is actually depressed with increasing the impurity scattering. From the information of the source materials (see Section 2), impurity atoms of Fe, Al, Ga, V are likely to incorporate into the lattice of K$_{2}$Cr$_{3}$As$_{3}$. Note that a magnetic ion, like Fe, is not necessarily magnetic if it is placed in a lattice environment like the Cr-based alloys.\cite{Cr} We argue that these impurities probably serve a \emph{nonmagnetic} scattering, since magnetic impurities would generally induce a resistivity minimum in the $R(T)$ curve (so-called Kondo effect) which is absent here. Therefore, qualitatively speaking, our result violates the Anderson¡¯s theorem,\cite{anderson,rmp} and suggests an unconventional superconductivity in which the superconducting order parameter changes sign around the Fermi surface.

The nonmagnetic scattering effect for a non-$s$-wave superconductor can also be quantitatively described by a generalized AG pair-breaking model,\cite{larkin,millis,radtke,Ru} which yields a similar AG equation\cite{AG} that determines $T_\mathrm{c}$,

\begin{equation}
\mathrm{ln}\left(\frac{T_{\mathrm{c0}}}{T_{\mathrm{c}}}\right)=\psi\left(\frac{1}{2}+g\frac{T_{\mathrm{c0}}}{T_{\mathrm{c}}}\right)-\psi\left(\frac{1}{2}\right),
\end{equation}

where $\psi$ is the digamma function, $g=\hbar/(4\pi \tau k_{\mathrm{B}}T_{\mathrm{c0}})$ is a measure of the pair breaking in which $\tau$ denotes the mean free time due to impurity scattering and, $T_{\mathrm{c0}}$ is the superconducting transition temperature in the clean limit. $\tau$ correlates with the electron mean free path by the formula $\ell=v_{\mathrm{F}}\tau$ ($v_{\mathrm{F}}$ refers to the Fermi velocity). According to the BCS result,\cite{tinkham} the superconducting coherence length for a clean-limit sample at zero temperature is defined by $\xi_{0}=\hbar v_\mathrm{F}/\pi \Delta_{0}$, where $\Delta_{0}=1.76 k_{\mathrm{B}}T_{\mathrm{c0}}$. Thus, a simple relation $g=0.44\xi_{0}/\ell$ can be derived.

The electron mean free path $\ell$ at low temperatures ($T\leq T_{\mathrm{c}}$) can be estimated from $\rho_{0}$ and the Sommerfeld coefficient $\gamma_{\mathrm{N}}$ using the Drude model.\cite{ssp} Since only the residual resistivity along the $c$ axis, $\rho_{0}^{\parallel}$, is available, we only deal with the electron mean free path by the $c$-direction transport measurement, i.e. $\ell_{\mathrm{tr}}^{\parallel}=v_{\mathrm{F}}^{\parallel}\times\tau=[(r_{s}/a_{0})^{2}/\rho_{0}^{\parallel}]\times 9.2$ nm, where $\rho_{0}^{\parallel}$ is in $\mu \Omega$ cm, $r_{s}$ is the electron density, and $a_{0}=\hbar/me^{2}$ is the Bohr radius. The $(r_{s}/a_{0})^2$ value can be obtained with the measured $\gamma_{\mathrm{N}}$, which equals to $0.07098 Z(r_{s}/a_{0})^{2}$ mJ/(mol K$^{2})$, where $Z$ counts the number of conduction electrons ($Z$=11 by assuming that all the Cr 3$d$ electrons conduct). $\ell_{\mathrm{tr}}^{\parallel}$ (in nanometers) is thus formulated by $\ell_{\mathrm{tr}}^{\parallel}=129.6\gamma_{\mathrm{N}}/(Z\rho_{0}^{\parallel})$. It was found that, as expected, the $\gamma_{\mathrm{N}}$ value,\cite{bao,canfield1} hardly changes with $\rho_{0}$. One may easily calculate the $\ell_{\mathrm{tr}}^{\parallel}$ values using the measured $\rho_{0}^{\parallel}$, which shows $\ell_{\mathrm{tr}}^{\parallel}\approx$ 920 nm for Sample \#1 and $\ell_{\mathrm{tr}}^{\parallel}\approx$ 41 nm for Sample \#4. If assuming that any impurity atom located at the Cr sites in the infinite [(Cr$_3$As$_3$)$^{2-}$]$_{\infty}$ chains is responsible for the potential scattering, one may estimate that the $\ell_{\mathrm{tr}}^{\parallel}$ values correspond to an impurity concentration of 0.008\% and 0.17\% for Samples \#1 and \#4 respectively. These deduced impurity concentrations roughly agree with the those of the Cr-source materials.

Taking $\xi_{0}^{\parallel}\sim$ 3.5 nm,\cite{zzw} we plot the reduced superconducting transition temperature, $T_{\mathrm{c}}/T_{\mathrm{c0}}$ (with $T_{\mathrm{c0}}$=6.20 K), in figure \ref{fig3} as a function of $\xi_{0}^{\parallel}/l_{\mathrm{tr}}^{\parallel}$. The data points basically satisfies the AG formula above. Note that the $\xi_{0}^{\parallel}/l_{\mathrm{tr}}^{\parallel}$ value for the dirtiest sample (\#4) is only 0.085$\pm0.042$, which unambiguously places the K$_{2}$Cr$_{3}$As$_{3}$ superconductors basically in the clean limit. If the impurity-scattering effect fully follows the AG equation, $T_{\mathrm{c}}$ would become zero for $\xi_{0}^{\parallel}/l_{\mathrm{tr}}^{\parallel}\geq$0.32. This means that the critical impurity concentration to kill superconductivity would be from 0.43\% to 1.27\%, which agrees with the strong pair-breaking scenario for most systems.\cite{rmp} Nevertheless, we should remark here that this extrapolation needs to be examined by further experiments. One notes that the inelastic scattering effect could avoid a strong suppression of $T_{\mathrm{c}}$.\cite{radtke} Furthermore, correlation effect\cite{correlation} as well as spin-orbit locking\cite{fu} may also protect superconductivity against potential scattering.

Compared with K$_{2}$Cr$_{3}$As$_{3}$, the sister superconductor Cs$_{2}$Cr$_{3}$As$_{3}$ has an obviously lower $T_{\mathrm{c}}$ and much lower $H_{\mathrm{c2}}(0)$.\cite{Cs233} The low $H_{\mathrm{c2}}(0)$ value (65 kOe, as compared with 370 kOe in K$_{2}$Cr$_{3}$As$_{3}$\cite{canfield2,zzw}) means a remarkably larger $\xi_{0}^{\parallel}$, which effectively increases the $\xi_{0}^{\parallel}/l_{\mathrm{tr}}^{\parallel}$ value. Therefore, one may expect that Cs$_{2}$Cr$_{3}$As$_{3}$ would be more sensitive to impurity scattering. Indeed, in contrast to the sharp superconducting transition with a full superconducting volume fraction in K$_{2}$Cr$_{3}$As$_{3}$,\cite{bao} the Cs$_{2}$Cr$_{3}$As$_{3}$ polycrystalline sample shows a much broader superconducting transition with a reduced superconducting volume fraction.\cite{Cs233} Further investigations on the impurity effect in Cs$_{2}$Cr$_{3}$As$_{3}$ may consolidate this point.

\section{Conclusion}\label{sec:4}

In summary, the impurity-scattering effect on superconductivity in K$_{2}$Cr$_{3}$As$_{3}$ is studied by careful resistivity measurements for various K$_{2}$Cr$_{3}$As$_{3}$ crystals with different residual resistivity. The results indicate that the apparently weak sensitivity of $T_{\mathrm{c}}$ to the impurity scattering is due to the fact that the samples are in very clean limit. Since the impurity scattering is basically nonmagnetic, and the $T_{\mathrm{c}}$ depression quantitatively satisfies the generalized AG theory, the result clearly supports a non-$s$-wave superconductivity with the superconducting order parameter changes sign around the Fermi surface. Therefore, the discrepancy of triplet pairing with regard to nonmagnetic impurity scattering in K$_{2}$Cr$_{3}$As$_{3}$ may be eliminated.

\begin{acknowledgments}
This work was supported by the Natural Science Foundation of China (No. 11190023), the National Basic Research Program (Nos. 2011CBA00103 and 2012CB927404), and the Fundamental Research Funds for the Central Universities of China.
\end{acknowledgments}

\end{document}